\documentclass[prd,preprint,superscriptaddress,amsmath,amssymb,nofootinbib]{revtex4}
\usepackage{graphicx}% Include figure files
\usepackage{dcolumn}% Align table columns on decimal point
\usepackage{bm}% bold math
\usepackage{amssymb}
\usepackage{amsmath}
\usepackage{epsfig}    
\usepackage{color}
\usepackage{slashed}
\usepackage{hhline}
\def\be{\begin{equation}}
\def\ee{\end{equation}}
\newcommand{\bea}{\begin{eqnarray}}
\newcommand{\eea}{\end{eqnarray}}
\newcommand{\nn}{\nonumber}

\begin{document}

\title{A radiative neutrino mass model with leptoquarks \\
under non-holomorphic modular $A_4$ symmetry}

\author{Takaaki Nomura}
\email{nomura@scu.edu.cn}
\affiliation{College of Physics, Sichuan University, Chengdu 610065, China}

\author{Hiroshi Okada}
\email{hiroshi3okada@htu.edu.cn}
\affiliation{Department of Physics, Henan Normal University, Xinxiang 453007, China}

\author{Xing-Yu Wang}
%星宇 汪
\email{xingyuwang@stu.scu.edu.cn}
\affiliation{College of Physics, Sichuan University, Chengdu 610065, China}

\date{\today}

\begin{abstract}
We investigate a radiative seesaw model with two leptoquarks under non-holomorphic modular $A_4$ symmetry. 
The leptons and quarks belong to non-trivial representations of the modular $A_4$ and the structures of their mass matrices are restricted.
Neutrino masses are generated at one-loop level via leptoquark inside loop diagram where structures of relevant Yukawa interactions are determined by the modular $A_4$ symmetry.
We scan the free parameters in the model and try to fit all the observed data for both lepton and quark sectors.
For allowed parameters, we show some predictions regarding neutrino observables such as sum of neutrino mass and neutrinoless double beta decay.
 \end{abstract}
\maketitle

\section{Introduction}

Understanding nature of flavor structure is one of the important purposes in particle physics as the standard model (SM) does not have criterion to restrict it.
Thus we expect that the structure of flavor in both lepton and quark sectors would be determined in beyond the standard model (BSM) framework.
A new symmetry is one of the good candidates controlling flavor. 
In fact there have been many approaches to build BSM applying a symmetry to describe flavor.

A modular flavor group provides attractive symmetry to describe flavor in which Yukawa couplings are written by a modular form transforming as a non-trivial representation of non-Abelian discrete group~\cite{Feruglio:2017spp}.
Application of modular flavor symmetry helps us to reduce the number of free parameters and extra scalar fields like flavons, providing more predictions regarding flavor physics.
Recently a non-holomorphic modular symmetry framework was constructed~\cite{Qu:2024rns} where we can successfully apply the modular group to non-supersymmetric models. 
Thus possibility of model building using this symmetry is enlarged and several applications have been considered, especially in models of neutrino mass generations~\cite{Ding:2024inn, Li:2024svh, Nomura:2024atp, Nomura:2024vzw,Nomura:2024nwh,Okada:2025jjo,Kobayashi:2025hnc,Loualidi:2025tgw}.
It is interesting to apply the non-holomorphic modular symmetry framework to both quark and lepton sectors including neutrino mass generation mechanism.

In this paper, we consider a radiative seesaw model with two scalar leptoquarks to generate neutrino masses that can naturally connect quark and lepton sectors~\cite{AristizabalSierra:2007nf,Kohda:2012sr,Cheung:2016fjo,Nomura:2016ezz,Nomura:2016ask,Cheung:2017efc,Dorsner:2017wwn,Babu:2020hun}.
Then a non-holomorphic modular $A_4$ symmetry is applied to restrict Yukawa interactions for quarks and leptons.
As a minimal model we introduce color triplet scalar leptoquarks  $\eta$ and  $S$ that are respectively $SU(2)_L$ doublet and singlet with hypercharge $1/6$ and $1/3$. 
%with which 
As a result, we can obtain one-loop diagram generating neutrino masses in which these two leptoquarks propagate.
Introduction of leptoquarks is also phenomenologically interesting since it provides rich phenomena related to flavor physics and collider signatures~\cite{Queiroz:2014pra, Dorsner:2016wpm,Assad:2017iib}. 
In particular, $S$ and/or combination of some leptoquarks can be applied to explain $R_D$ anomaly in B meson decay and to enhance muon anomalous magnetic moment~\cite{Sakaki:2013bfa, Bauer:2015knc,Becirevic:2016yqi,Chen:2016dip, Chen:2017hir,Cai:2017wry,Nomura:2021oeu,Parashar:2022wrd}.
In this work, we focus on neutrino mass generations under a non-holomorphic $A_4$ modular symmetry and try to fit all the observed data in both lepton and quark sectors.
Then some predictions in neutrino observables are explored.

This paper is organized as follows.
In Sec.~II, we summarize our model and provide mass matrices for charged leptons and quarks, and show our formula for neutrino mass matrix generated at one-loop level. 
In Sec.~III, we perform $\chi$ square numerical analysis including both quark and lepton sectors, and show some predictions in neutrino observables for normal and inverted hierarchies.
We provide summary and discussion in Sec.~IV.

%%%%%%%%%%%%%%%%%%%%%%%%%%%%%%%%%%%%%%%%%%%%%%

\section{model setup}

\begin{table}[t!]
\begin{tabular}{|c||c|c|c|c|c||c|c|c|}\hline\hline  
& ~$ L_L$~ & ~$ \ell_R$~ & ~$Q_L$~  & ~$d_R$~  & ~$u_R$~ & ~$ H$ ~&~ $\eta$~ &~ $S$~  \\\hline\hline 
%%%
$SU(3)$ & $\bm{1}$ & $\bm{1}$ & $\bm{3}$ & $\bm{3}$ & $\bm{3}$ & $\bm{1}$ & $\bm{3}$ & $\bm{\bar{3}}$ \\ \hline
$SU(2)_L$   & $\bm{2}$  & $\bm{1}$  & $\bm{2}$ & $\bm{1}$ & $\bm{1}$   & $\bm{2}$  & $\bm{2}$ & $\bm{1}$     \\\hline 
$U(1)_Y$    & $\frac12$  & $-1$ & $\frac16$  & $-\frac13$  & $\frac23$  & $\frac12$ & $\frac16$  & $\frac13$    \\\hline
$A_4$   & $\bm{3}$  & $\bm{3}$ & $\bm{3}$ & $\bm{3}$ & $\{\bm{1}\}$ & $\bm{1}$ & $\bm{1}$  & $\bm{1}$         \\\hline 
$-k_I$    & $0$  & $0$ & $0$ & $0$  & $-6$  & $0$ & $0$   & $0$      \\\hline
\end{tabular}
\caption{Charge assignments of the SM leptons $\{L_L, \ell_R \}$, quarks $\{Q_L, u_R, d_R \}$ and scalar leptoquarks $\{\eta, S \}$
under $SU(3) \otimes SU(2)_L\otimes U(1)_Y \otimes A_4$ where $-k_I$ is the number of modular weight and {$\{ \bm{1} \} =\{1, 1', 1''\}$} indicates assignment of $A_4$ singlets.
}\label{tab:1}
\end{table}

In this section, we summarize the model and formulas for phenomenological analysis.
The model is based on $G_{SM} \otimes A_4$ symmetry where $G_{SM}$ is the SM gauge symmetry and $A_4$ corresponds to non-holomorphic modular symmetry.
We assign $A_4$ triplet and modular weight $0$ to the SM leptons. 
For the SM quarks, $Q_L$ and $d_R$ are $A_4$ triplets with modular weight $0$ while we assign $A_4$ singlets $\{1, 1' , 1'' \}$ with modular weight $-6$ to three generations of $u_R$~\footnote{Note that we can not fit quark masses and mixings in universal assignment with modular weight 0 and triplet $\bm{3}$ for $u_R$, which is confirmed by our numerical analysis. Instead, the assignment of weight $-6$ and $\{\bm{1} \}$ to $u_R$ is our minimal case to realize observed values.}.
Then we introduce two scalar leptoquarks $\eta = (\eta_{\frac23}, \eta_{-\frac13})$~\footnote {The subscripts on the components indicate electric charge.} and $S$ that are assigned as $\{\bm{3}, \bm{2}, 1/6 \}$ and $\{\bar{\bm{3}}, \bm{1}, 1/3 \}$ under the SM gauge symmetry $\{SU(3), SU(2)_L, U(1)_Y \}$ 
and $A_4$ singlets with modular weight $0$.
The SM Higgs field is assigned as $A_4$ trivial singlet with modular weight $0$.

Invariant Yukawa interactions are written by
\begin{align}
- \mathcal{L}_Y = & f_1 [Y_{\bm{1}}^{(0)} \otimes (\overline{Q^c_L} S L_L )_{ \bm{1}}] + f_S [Y_{\bm{3}}^{(0)} \otimes (\overline{Q^c_L} S L_L )_{\bm{3}_S}] + f_A [Y_{\bm{3}}^{(0)} \otimes (\overline{Q^c_L} S L_L )_{\bm{3}_A}] \nonumber \\
& + g_1 [Y_{\bm{1}}^{(0)} \otimes (\overline{d_R} \eta L_L )_{ \bm{1}}] + g_S [Y_{\bm{3}}^{(0)} \otimes (\overline{d_R} \eta L_L )_{\bm{3}_S}] + f_A [Y_{\bm{3}}^{(0)} \otimes (\overline{d_R} \eta L_L )_{\bm{3}_A}] \nonumber \\
& + y_1^d [Y_{\bm{1}}^{(0)} \otimes (\overline{Q_L} H d_R )_{ \bm{1}}] + y^d_{3} [Y_{\bm{3}}^{(0)} \otimes (\overline{Q_L} H d_R )_{\bm{3}_S}] + y^d_{3'} [Y_{\bm{3}}^{(0)} \otimes (\overline{Q_L} H d_R)_{\bm{3}_A}] \nonumber \\
&+ y_1^\ell [Y_{\bm{1}}^{(0)} \otimes (\overline{L_L} H \ell_R )_{ \bm{1}}] + y^\ell_{3} [Y_{\bm{3}}^{(0)} \otimes (\overline{L_L} H \ell_R )_{\bm{3}_S}] + y^\ell_{3'} [Y_{\bm{3}}^{(0)} \otimes (\overline{L_L} H \ell_R)_{\bm{3}_A}] \nonumber \\
& + \sum_{A=1,2} \left[ \alpha^A_u [Y_{3^A}^{(6)} \overline{Q_L} \tilde{H} ]_{\bm{1}} u_{R_{\bm 1}}  + \beta^A_u [Y_{3^A}^{(6)} \overline{Q_L} \tilde{H} ]_{\bm{1''}} u_{R_{\bm{1'}}} + \gamma^A_u [Y_{3^A}^{(6)} \overline{Q_L} \tilde{H} ]_{\bm{1'}} u_{R_{\bm{1''}}} \right] \nonumber \\ 
& + \sum_{A=1,2} \left[ a^A_u \overline{u^c_{R_{\bm{1}}}} [Y_{3^A}^{(6)} S \ell_R ]_{\bm{1}}  + b^A_u \overline{u^c_{R_{\bm{1'}}}} [Y_{3^A}^{(6)} S \ell_R ]_{\bm{1''}}  + c^A_u \overline{u^c_{R_{\bm{1''}}}} [Y_{3^A}^{(6)} S \ell_R ]_{\bm{1'}} \right], \label{eq:Yukawa}
\end{align}
where subscripts $\{\bm{1},\bm{3}_S, \bm{3}_A \}$ indicate $A_4$ representation, subscript $S(A)$ corresponds to a triplet from symmetric(antisymmetric) combination of two triplets and 
$\tilde H = i \sigma_2 H^*$ with $\sigma_2$ being the second Pauli matrix. 
The $Y^{(k_I)}_{\bm{R}}$ expresses a modular form in non-holomorphic framework where $k_I$ and $\bm{R}$ indicate modular weight and representation under $A_4$ respectively.
We write $Y^{(0)}_{\bm{3}} = (y_1,y_2,y_3)$ and $Y^{(6)}_{\bm{3}^{1[2]}} = (\tilde{y}_1, \tilde{y}_2, \tilde{y}_3) [(\tilde{y}'_1, \tilde{y}'_2, \tilde{y}'_3)]$ where the components are functions of modulus $\tau$ written by Maa{\ss} forms~\cite{Qu:2024rns}.
The interaction terms are invariant under modular transformation when it is $A_4$ singlet and the sum of modular weight is zero~\cite{Feruglio:2017spp, Qu:2024rns}.

The scalar potential is given by
\begin{align}
V = & - \mu_H^2 |H|^2 + \mu_\eta |\eta|^2 + \mu_S |S|^2 + (\mu H^\dagger \eta S + h.c.) \nonumber \\
& + \lambda_H |H|^4 + \lambda_\eta |\eta|^4 + \lambda_S |S|^4 + \lambda_{H \eta} |H|^2 |\eta|^2  + \lambda_{H S} |H|^2 |S|^2 +  \lambda_{ \eta S} |\eta|^2 |S|^2, 
\end{align}
where all the parameters are assumed to be real.

\subsection{Scalar sector}

The SM Higgs develops a vacuum expectation value (VEV), $\langle H \rangle = v/\sqrt{2}$, as in the SM breaking electroweak symmetry.
In this model Higgs boson is just the same as the SM one since the leptoquarks do not develop VEVs.
The leptoquarks $\eta_{-\frac13}$ and $S$ mix via the term $H^\dagger \eta S + h.c.$ after electroweak symmetry breaking.
Here we write mass eigenstate, $\{\rho_{\frac13}, \chi_{\frac13} \}$, as 
\begin{equation}
\begin{pmatrix} \eta_{\frac13} \\ S \end{pmatrix} = \begin{pmatrix} \cos \alpha & \sin \alpha \\ - \sin \alpha & \cos \alpha \end{pmatrix} \begin{pmatrix} \rho_{\frac13} \\ \chi_{\frac13} \end{pmatrix},
\end{equation}
where we consider mixing angle $\alpha$ as a free parameter. 
The mass eigenvalues are also written by $m_\rho$ and $m_\chi$ that are also taken as free parameters in numerical analysis below.

\subsection{Quark mass and mixing}

The quark mass matrices are obtained after electroweak symmetry breaking.
From the terms in third line of Eq.~\eqref{eq:Yukawa}, the down-type quark mass matrix is written by
\begin{equation}
(M_d)_{LR} = \frac{v}{\sqrt{2}} 
\begin{pmatrix}
y_1^d + 2 y_3^d y_1 & (-y_3^d + y_{3'}^d) y_3 & - (y_3^d + y^d_{3'}) y_2 \\
(-y^d_3 + y^d_{3'}) y_2 & y_1^d - (y_3^d + y^d_{3'}) y_1 & 2 y^d_3 y_3 \\
-(y^d_3 + y^d_{3'}) y_3 & 2 y_3^d y_2 & y^d_1 + (-y^d_3 + y^d_{3'}) y_1 
\end{pmatrix}.
\end{equation}
From the terms in 5th line of Eq.~\eqref{eq:Yukawa}, the up-type quark mass matrix is obtained as
\begin{equation}
(M_u)_{LR} = \frac{v}{\sqrt{2}} \left[ 
\begin{pmatrix} \tilde{y}_1 & \tilde{y}_3 & \tilde{y}_2 \\ \tilde{y}_2 & \tilde{y}_1 & \tilde{y}_3 \\ \tilde{y}_3 & \tilde{y}_2 & \tilde{y}_1 \end{pmatrix} +
\begin{pmatrix} \tilde{y}'_1 & \tilde{y}'_3 & \tilde{y}'_2 \\ \tilde{y}'_2 & \tilde{y}'_1 & \tilde{y}'_3 \\ \tilde{y}'_3 & \tilde{y}'_2 & \tilde{y}'_1 \end{pmatrix}
\begin{pmatrix} \tilde{\alpha} & 0 & 0 \\ 0 & \tilde{\beta} & 0 \\ 0 & 0 & \tilde{\gamma} \end{pmatrix}
\right] \begin{pmatrix} \alpha_u^1 & 0 & 0 \\ 0 & \beta_u^1 & 0 \\ 0 & 0 & \gamma_u^1 \end{pmatrix}
\end{equation}
where $\tilde{\alpha} = \alpha^2_u/\alpha^1_u$, $\tilde{\beta} = \beta^2_u/\beta^1_u$ and $\tilde{\gamma} = \gamma^2_u/\gamma^1_u$.
The quark mass matrices are diagonalized by bi-unitary transformations diag.$(m_d,m_s,m_b)\equiv V_{dL}^\dag M_d V_{dR}$ and diag.$(m_u,m_c,m_t)\equiv V_{uL}^\dag M_u V_{uR}$.
Then, the CKM mixing matrix is defined by $V_{\rm CKM}=V^\dag_{uL} V_{dL}$ as in the SM.
The free parameters $\{\alpha_{u}^1,\beta_{u}^1,\gamma_{u}^1 \}$ and $\{y_1^d, y_3^d, y_{3'}^d \}$ are determined to fit the quark masses in use of the relations,
\begin{align}
&{\rm Tr}[M^\dag_{u,d} {M_{u,d}}] = |m_{u,d}|^2 + |m_{c,s}|^2 + |m_{t,b}|^2,\\
&{\rm Det}[M^\dag_{u,d} {M_{u,d}}] = |m_{u,d}|^2  |m_{c,s}|^2  |m_{t,b}|^2,\\
&({\rm Tr}[M^\dag_{u,d} {M_{u,d}}])^2 -{\rm Tr}[(M^\dag_{u,d} {M_{u,d}})^2] =2( |m_{u,d}|^2  |m_{c,s}|^2 + |m_{c,s}|^2  |m_{t,b}|^2+ |m_{u,d}|^2  |m_{t,b}|^2 ),
\end{align}
where we adopt the values of quark masses in PDG~\cite{ParticleDataGroup:2024cfk}.

\subsection{Charged lepton mass}

The charged lepton mass matrix is obtained from the terms in fourth line of Eq.~\eqref{eq:Yukawa}. 
After electroweak symmetry breaking, the mass matrix is written by
\begin{equation}
(M_\ell)_{LR} = \frac{v}{\sqrt{2}} 
\begin{pmatrix}
y_1^\ell + 2 y_3^\ell y_1 & (-y_3^\ell + y_{3'}^\ell) y_3 & - (y_3^\ell + y^\ell_{3'}) y_2 \\
(-y^\ell_3 + y^\ell_{3'}) y_2 & y_1^\ell - (y_3^\ell + y^\ell_{3'}) y_1 & 2 y^\ell_3 y_3 \\
-(y^\ell_3 + y^\ell_{3'}) y_3 & 2 y_3^\ell y_2 & y^\ell_1 + (-y^\ell_3 + y^\ell_{3'}) y_1 
\end{pmatrix}.
\end{equation}
The mass eigenvalues for the charged-leptons can be obtained by bi-unitary transformation diag.$(m_e,m_\mu,m_\tau)\equiv V_L^\dag M_\ell V_R$ as in the SM case.
We thus find $V_L^\dag M_\ell M_\ell^\dag V_L ={\rm diag.}(|m_e|^2,|m_\mu|^2,|m_\tau|^2)$.
%%%
The three observed charged-lepton masses should satisfy the following relations:
\begin{align}
&{\rm Tr}[M_\ell M_\ell^\dag] = |m_e|^2 + |m_\mu|^2 + |m_\tau|^2,\quad
 {\rm Det}[M_\ell M_\ell^\dag] = |m_e|^2  |m_\mu|^2  |m_\tau|^2,\nn\\
&({\rm Tr}[M_\ell M_\ell^\dag])^2 -{\rm Tr}[(M_\ell M_\ell^\dag)^2] =2( |m_e|^2  |m_\mu|^2 + |m_\mu|^2  |m_\tau|^2+ |m_e|^2  |m_\tau|^2 ).\label{eq:l-cond}
\end{align}
We adopt these relations to fix free parameters $\{y_1^\ell, y^\ell_3, y^\ell_{3'} \}$ by fitting charged lepton masses to observed values, 
where the values of lepton masses are referred to PDG~\cite{ParticleDataGroup:2024cfk}.

\subsection{Neutrino mass and mixing}

%%%%%%%%%%%%%%%%%%%%%%%%%%%%%%%%%%%%%%%%%%%%%%%%%%%%%%%%%%%%%%%%%%%%%%%%%%%%%%%%%%%%
%%%%%%%%%%%%%%%%%%%
\begin{figure}[tb!]\begin{center}
\includegraphics[width=80mm]{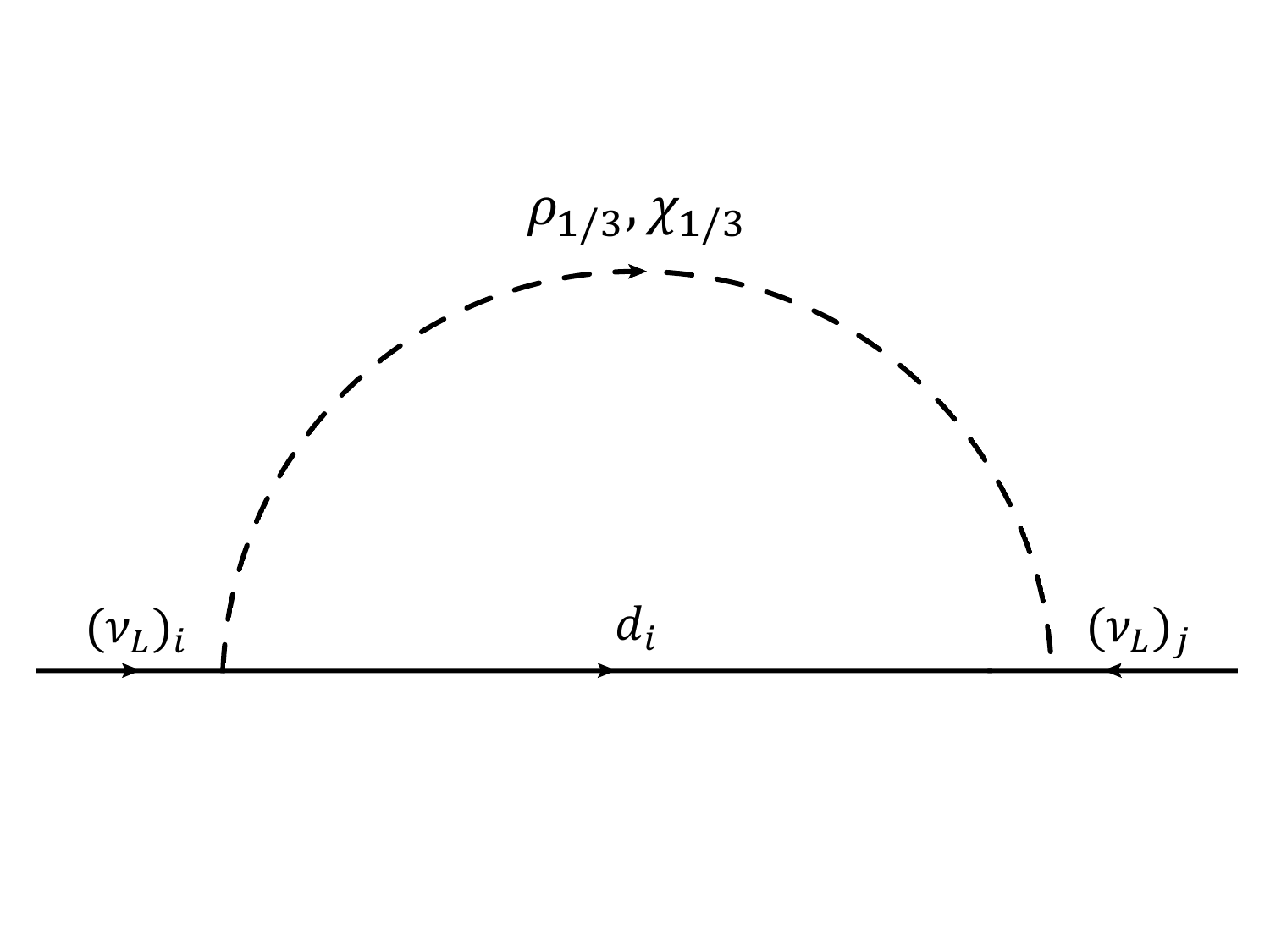} 
\caption{The one-loop diagram generating neutrino masses.}   
\label{fig:diagram}\end{center}\end{figure}
%%%%%%%%%%%%%%%%%%%
%%%%%%%%%%%%%%%%%%%%%%%%%%%%%%%%%%%%%%%%%%%%%%%%%%%%%%%%%%%%%%%%%%%%%%%%%%%%%%%%%%%%

In the model, neutrino mass is radiatively generated at one-loop level.
The relevant interactions in generating neutrino mass is obtained from terms in the first and second lines of Eq.~\eqref{eq:Yukawa}.
They are written by
\begin{align}
& -\mathcal{L}_Y \supset
{ \overline{Q^c_{L_i}} f_1 F_{ij} L_j S + \overline{d_{R_i}} g_1 G_{ij} L_j \eta + h.c., } \\
& 
F_{ij} = {
\begin{pmatrix} 1 + 2  f_S/f_1 y_1 & (-f_S + f_A)/f_1 y_3 & - (f_S + f_A)/f_1 y_2 \\ -(f_S + f_A)/f_1 y_3 & 2 f_S/f_1 y_2 & 1 - (f_S - f_A)/f_1 y_1 \\ (-f_S + f_A)/f_1 y_2 & 1 - (f_S +f_A)/f_1 y_1 & 2 f_S/f_1 y_3
\end{pmatrix}, 
}
%\begin{pmatrix} f_1 + 2 f_S y_1 & (-f_S + f_A) y_3 & - (f_S + f_A) y_2 \\ -(f_S + f_A) y_3 & 2 f_S y_2 & f_1 - (f_S - f_A) y_1 \\ (-f_S + f_A) y_2 & f_1 - (f_S +f_A) y_1 & 2 f_S y_3
%\end{pmatrix}, 
\\
& G_{ij} = 
{
\begin{pmatrix} 1 + 2 g_S/g_1 y_1 & (-g_S + g_A)/g_1 y_3 & - (g_S + g_A)/g_1 y_2 \\ (-g_S + g_A)/g_1 y_2 & 1 - (g_S + g_A)/g_1 y_1 & 2 g_S/g_1 y_3 \\ -(g_S + g_A)/g_1 y_3 & 2 g_S/g_1 y_2 & 1 + (-g_S +g_A)/g_1 y_1
\end{pmatrix}.
}
%\begin{pmatrix} g_1 + 2 g_S y_1 & (-g_S + g_A) y_3 & - (g_S + g_A) y_2 \\ (-g_S + g_A) y_2 & g_1 - (g_S + g_A) y_1 & 2 g_S y_3 \\ -(g_S + g_A) y_3 & 2 g_S y_2 & g_1 + (-g_S +g_A) y_1
%\end{pmatrix}.
\end{align}
Note that quark and leptoquarks have to be replaced by their mass eigenstates in the above form. 
We thus rewrite the Yukawa interactions in mass eigenstate such that
\begin{align}
& \mathcal{L}_Y \supset f_1  \overline{d^{m \, c}_{L_a}} \tilde{F}_{ai}\nu_{L_i} (-s_\alpha \rho_{\frac13} + c_\alpha \chi_{\frac13}) 
+ g_1 \overline{d^m_{R_a}} \tilde{G}_{ai}  \nu_{L_i} (c_\alpha \rho_{-\frac13} + s_\alpha \chi_{- \frac13}), \\
& \tilde{F}_{ai} \equiv (V^T_{uL})_{aj} F_{ji}, \quad \tilde{G}_{ai} \equiv (V^\dagger_{dR})_{aj} G_{ji}, 
\end{align}
where $c_{\alpha}(s_\alpha) = \cos \alpha (\sin \alpha)$, $d^m$ indicates mass eigenstate of down type quark, and we only pick up terms relevant to neutrino mass generation.
The neutrino masses are generated via one-loop diagram in Fig.~\ref{fig:diagram} and analytic formula is given by~\cite{Cheung:2016fjo}
\begin{align}
({\cal M}_\nu)_{ij} &=f_1g_1 \frac{N_c s_\alpha c_\alpha}{2 (4 \pi)^2} \left( 1 - \frac{m_\rho^2}{m^2_\chi} \right)
\sum_{a=1}^3 \left[ (\tilde{F}^T)_{ja} m_{d_a} \tilde{G}_{a i}  + (\tilde{G}^T)_{j a} m_{d_a} \tilde{F}_{aj} \right] F_I (r_\rho, r_{d_a}), \label{eq:mnu} \\
F_I(r_1, r_2) &= \frac{r_1 (r_2 - 1) \ln(r_1) - r_2 (r_1 - 1) \ln(r_2)}{(r_1 - 1)(r_2 -1)(r_1 - r_2)}, 
\end{align}
where $r_f = m^2_f/m^2_{\chi}$.
For convenience in numerical analysis, the neutrino mass matrix is rewritten as ${\cal M}_\nu = \kappa \tilde {\cal M}_\nu$ where $\kappa$ {$\left(\equiv f_1 g_1\right) $ is a dimensionless} overall factor to control scale and $\tilde {\cal M}_\nu$  
{ has mass dimension one.}
We then diagonalize $\tilde {\cal M}_\nu$ by a unitary matrix $U_\nu$ as $U_\nu^T \tilde {\cal M}_\nu U_\nu =\tilde m_\nu$ with diagonal matrix $\tilde m_\nu = {\rm diag}[\tilde m_{\nu_1},\tilde m_{\nu_2},\tilde m_{\nu_3}]$.
Thus the Pontecorvo-Maki-Nakagawa-Sakata (PMNS) unitary matrix $U_{PMNS}$ is defined by $V_{L}^\dag U_\nu$ where $V_L$ is obtained in diagonalizing charged lepton mass matrix.
We then write the observed atmospheric mass squared difference by 
\begin{align}
&{\rm NH}:\ \Delta m^2_{atm}= |\kappa|^2 (\tilde m^2_{\nu_3} - \tilde m^2_{\nu_1}),\\
%%%
&{\rm IH}:\ \Delta m^2_{atm}= |\kappa|^2 (\tilde m^2_{\nu_2} - \tilde m^2_{\nu_3}),
\end{align}
where NH(IH) indicates the normal(inverted) hierarchy of neutrino masses.
The solar mass squared difference $\Delta m^2_{sol}$ is also written by
\begin{align}
\Delta m^2_{sol}= |\kappa|^2 (\tilde m^2_{\nu_2} - \tilde m^2_{\nu_1}).
\end{align} 
The sum of the neutrino mass is given by $\sum m_\nu\equiv \kappa(\tilde m_{\nu_1}+\tilde m_{\nu_2}+\tilde m_{\nu_3})$ that is constrained by some observations;
$\sum m_\nu\le120$ meV from the minimal standard cosmological model with CMB data~\cite{Vagnozzi:2017ovm, Planck:2018vyg}~\footnote{The upper bound can be weaker if the data are analyzed under extended cosmological models~\cite{ParticleDataGroup:2024cfk}.} 
and $\sum m_\nu\le72$ meV from the combined result from BAU data by DESI and CMB~\cite{DESI:2024mwx,DESI:2025ejh}.
For phases associated with PMNS matrix, we adopt the standard parametrization of the lepton mixing matrix $U \equiv V_{e_L}^\dag U_\nu$ in which the Majorana phases are defined by $[1,e^{i\alpha_{21}/2},e^{i\alpha_{31}/2}]$~\cite{Okada:2019uoy}. 
The Majorana phases can be estimated from PMNS matrix element as 
\begin{align}
\text{Re}[U^*_{e1} U_{e2}] = c_{12} s_{12} c_{13}^2 \cos \left( \frac{\alpha_{21}}{2} \right), \
 \text{Re}[U^*_{e1} U_{e3}] = c_{12} s_{13} c_{13} \cos \left( \frac{\alpha_{31}}{2} - \delta_{CP} \right), \\
 %%%
 \text{Im}[U^*_{e1} U_{e2}] = c_{12} s_{12} c_{13}^2 \sin \left( \frac{\alpha_{21}}{2} \right), \
 \text{Im}[U^*_{e1} U_{e3}] = c_{12} s_{13} c_{13} \sin \left( \frac{\alpha_{31}}{2} - \delta_{CP} \right),
\end{align}
where $\alpha_{21}/2,\ \frac{\alpha_{31}}{2} - \delta_{CP}$
are subtracted from $\pi$ if $\cos(\alpha_{21}/2)$ and $\cos \left( \frac{\alpha_{31}}{2} - \delta_{CP} \right)$ give negative values, and $c_{ij}(s_{ij})$ is abbreviation of $\cos \theta_{ij} (\sin \theta_{ij})$.
The effective neutrino mass for neutrinoless double beta decay,  $\langle m_{ee} \rangle$, is defined to be 
\begin{align}
\langle m_{ee}\rangle=\kappa|\tilde m_{\nu_1} \cos^2\theta_{12} \cos^2\theta_{13}+\tilde m_{\nu_2} \sin^2\theta_{12} \cos^2\theta_{13}e^{i\alpha_{2}}+\tilde m_{\nu_3} \sin^2\theta_{13}e^{i(\alpha_{3}-2\delta_{CP})}|.
\end{align}
It can be constrained by experiments searching for neutrinoless double beta decay where the strongest constraint is given by the current KamLAND-Zen data;
$\langle m_{ee}\rangle<(28-122)$ meV at 90 \% confidence level~\cite{KamLAND-Zen:2024eml}; here the range of the upper limit is due to adopted model of nuclear mass matrix elements.

\section{Numerical analysis}

In this section, we carry out numerical analysis to search for parameters that can fit the observed data of both quark and lepton sectors,
and show some predictions in neutrino sector.

\subsection{Quark sector}

In quark sector, we search for allowed regions estimating $\Delta \chi^2$ value to fit the sixteen parameters from reliable experimental data; six quark masses, nine CKM elements, and one quark CP phase, where we assume all observables satisfy Gaussian distribution. 
Then we randomly scan modulus $\tau$ in fundamental region and remain input free parameters as follows;
\begin{align}
\{|\tilde \alpha|, |\tilde \beta|, |\tilde \gamma|\} \in [10^{-5},10^2],
\end{align}
where the phases of these complex parameters are chosen randomly and the other free parameters in quark sector are fixed to fit the quark masses.
For the values of CKM elements, we adopt the observed values in PDG.

%%%%%%%%%%%%%%%%%%%%%%%%%%%%%%%%%%%%%%%%%%%%%%%%%%%%%%%%%%%%%%%%%%%%%%%%%%%%%%%%%%%%
%%%%%%%%%%%%%%%%%%%
\begin{figure}[tb!]\begin{center}
\includegraphics[width=75mm]{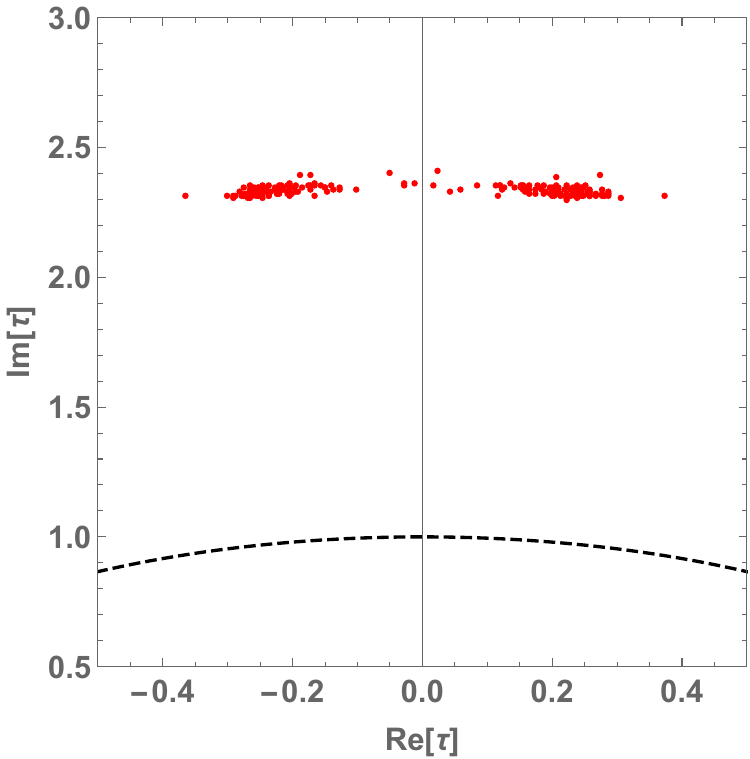} 
\caption{The region of modulus $\tau$ realizing observed values in quark sector within 3 $\sigma$ estimated from chi-square analysis.}   
\label{fig:1}\end{center}\end{figure}
%%%%%%%%%%%%%%%%%%%
%%%%%%%%%%%%%%%%%%%%%%%%%%%%%%%%%%%%%%%%%%%%%%%%%%%%%%%%%%%%%%%%%%%%%%%%%%%%%%%%%%%%

 Then we  estimate the $\Delta \chi^2$ value using the formula of
\begin{equation}
\Delta \chi^2 = \sum_{i}  \left( \frac{O_i^{\rm obs} - O_i^{\rm th}}{\delta O_i^{\rm exp}} \right)^2, \label{eq:chi-square}
\end{equation}
where $O_i^{\rm obs (th)}$ is observed (theoretically) obtained value of corresponding observables and $\delta O_i^{\rm exp}$ indicates the experimental error at $1\sigma$.
For the quark sector, we search for the parameters that realize $\Delta \chi^2$ values that give us within $3 \sigma$ standard deviation 
in use of sixteen reliable experimental data mentioned above.
In Fig.~\ref{fig:1}, we show the values of modulus $\tau$ that can fit the experimental data for quark sector.
We find that Im$[\tau] \sim 2.3-2.4$ is required to fit observed values in quark sector and $| {\rm Re}[\tau]|$ is also preferred to be below $0.4$.

\subsection{Lepton sector}

Here we discuss our numerical analysis for lepton sector.
Firstly we adopt the values of modulus $\tau$ and quark mixing matrices $V_{uL}$ and $V_{dR}$ that are obtained from the numerical analysis of quark sector fitting observables.
The values of quark mixing matrices are applied to calculate the neutrino mass matrix in Eq.~\eqref{eq:mnu}.
Then, for each set of $\{\tau, \, V_{uL}, \, V_{dR} \}$, we scan the free parameters in lepton sector such that
\begin{equation}
\{ |f_S|, |f_A|, |g_S|, |g_A|\} \in [10^{-5}, 1], \quad \{m_\rho, m_\chi \}  \in [10^3, 10^5] \ [{\rm GeV}], \quad s_\alpha \in [10^{-3}, 1],
\end{equation}
where we randomly choose the phases of complex parameters.
Note that $\kappa = f_1 g_1$ is used to fit the observed value of $\Delta m_{atm}^2$ by scaling neutrino mass matrix.
We then search for the parameter sets that can fit the neutrino oscillation data where we adopt the values in Nufit 6.0~\cite{Esteban:2024eli} to estimate $\Delta \chi^2$ value in lepton sector 
using Eq.~\eqref{eq:chi-square}. 
Note that here we calculate $\Delta \chi^2$ with $\{\sin^2 \theta_{12}, \sin^2 \theta_{13}, \Delta m^2_{\rm atm}, \Delta m^2_{sol} \}$ and $\sin^2 \theta_{23}$ is required to be within 3$\sigma$ level of 
Nufit 6.0 since the experimental error is deviated from Gaussian distribution.
We then calculate other observables such as Dirac CP phase and effective mass matrix for neutrinoless double beta decay.
In the following, we summarize our results for NH and IH cases separately.

 %%%%%%%%%%%%%%%%%%%%%%%%%%%%%%%%%%%%%%%%%%%%%%%%%%%%%%%%%%%%%%%%%%%%%%%%%%%%%%%%%%%%
%%%%%%%%%%%%%%%%%%%
\begin{figure}[tb!]\begin{center}
\includegraphics[width=50mm]{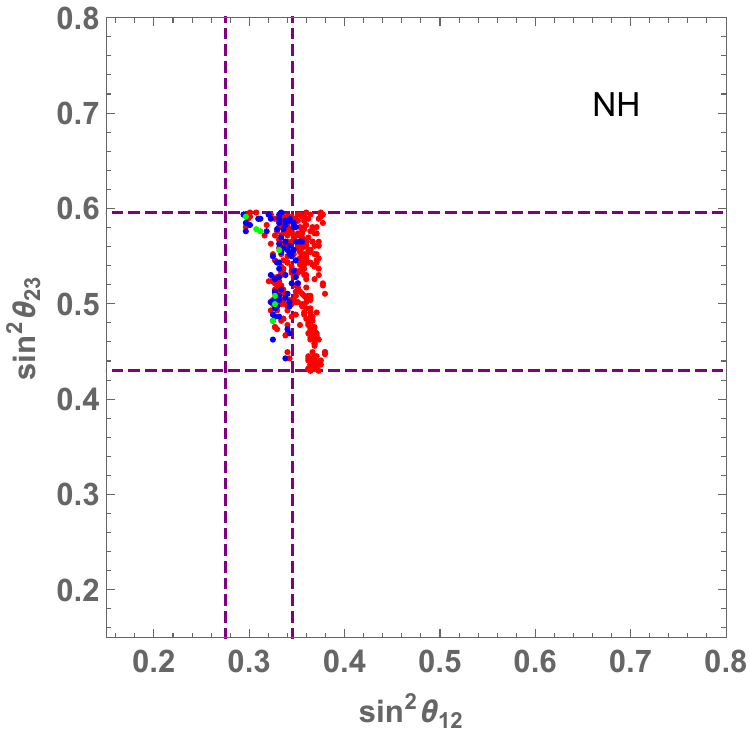} \
\includegraphics[width=53mm]{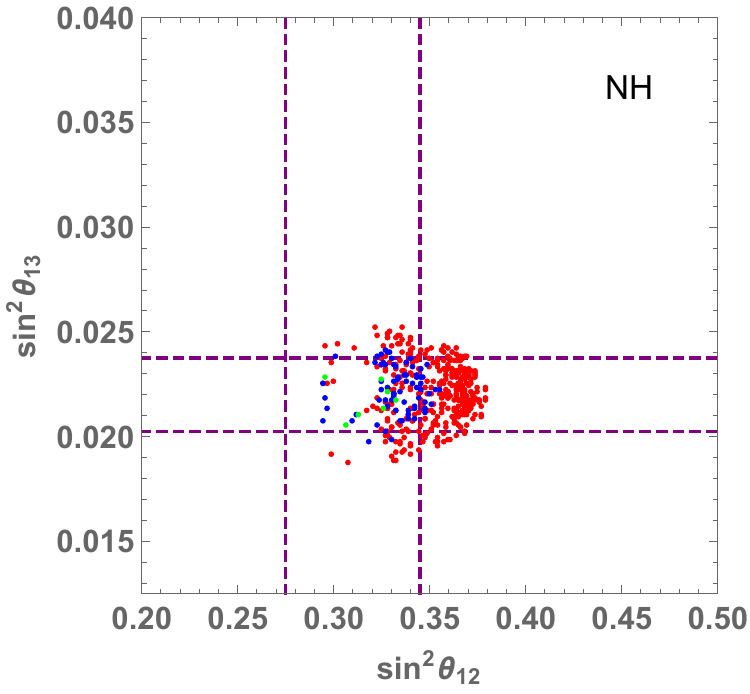} \
\includegraphics[width=53mm]{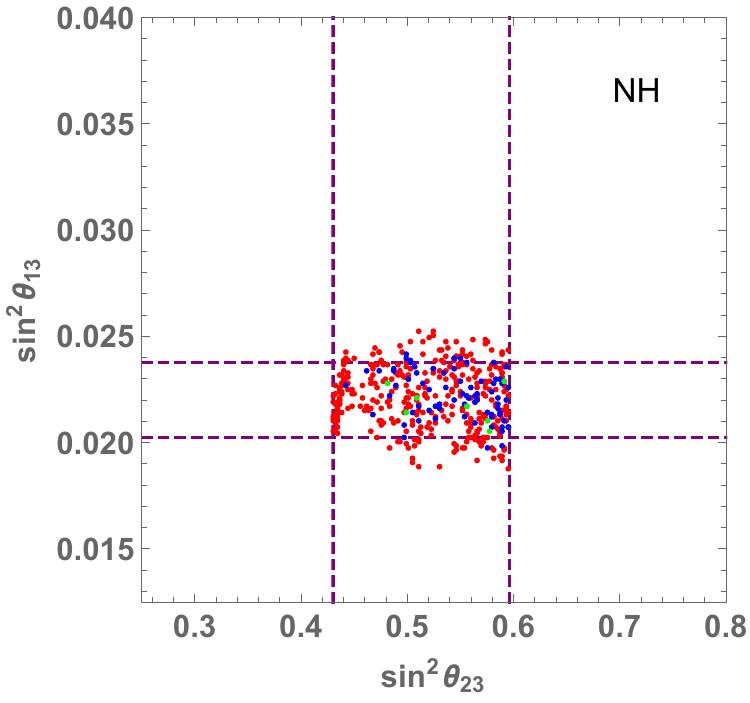} 
\caption{The neutrino mixing angles obtained from the allowed parameter points in NH case where red, blue and green colored points give the $\Delta \chi^2$ value for lepton sector providing confidence level within 5$\sigma$ to 3$\sigma$, 3$\sigma$ to 1$\sigma$ and within 1$\sigma$ (the color legend is common in the following figures) and the regions between purple dashed line correspond to $3 \sigma$ range of observables.  }   
\label{fig:2}\end{center}\end{figure}
%%%%%%%%%%%%%%%%%%%
%%%%%%%%%%%%%%%%%%%%%%%%%%%%%%%%%%%%%%%%%%%%%%%%%%%%%%%%%%%%%%%%%%%%%%%%%%%%%%%%%%%%

\subsubsection{NH case}

In Fig.~\ref{fig:2}, we show the neutrino mixing angles obtained from the allowed parameter points in NH case where red, blue and green colored points give the $\Delta \chi^2$ value for lepton sector providing confidence level within 5$\sigma$ to 3$\sigma$, 3$\sigma$ to 1$\sigma$ and within 1$\sigma$ (the color legend is common in the following figures) and the regions between purple dashed lines correspond to $3 \sigma$ range of the observed values.
We find that $\sin \theta_{12}$ value tends to be larger as $0.3 \lesssim \sin^2 \theta_{12}$ for $0.57 \lesssim \sin^2 \theta_{23}$ and $0.32 \lesssim \sin^2 \theta_{12}$ for entire range of $\sin^2 \theta_{23}$ within $3 \sigma$ level.
In contrast, the values of $\sin^2 \theta_{23}$ and $\sin^2 \theta_{13}$ span entire region within $3\sigma$. 
 %%%%%%%%%%%%%%%%%%%%%%%%%%%%%%%%%%%%%%%%%%%%%%%%%%%%%%%%%%%%%%%%%%%%%%%%%%%%%%%%%%%%
%%%%%%%%%%%%%%%%%%%
\begin{figure}[tb!]\begin{center}
\includegraphics[width=65mm]{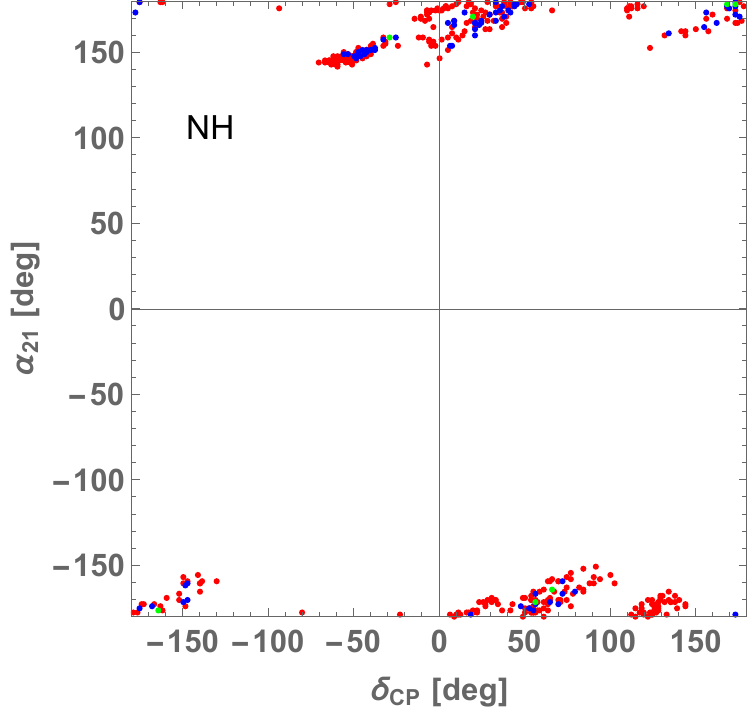} \
\includegraphics[width=65mm]{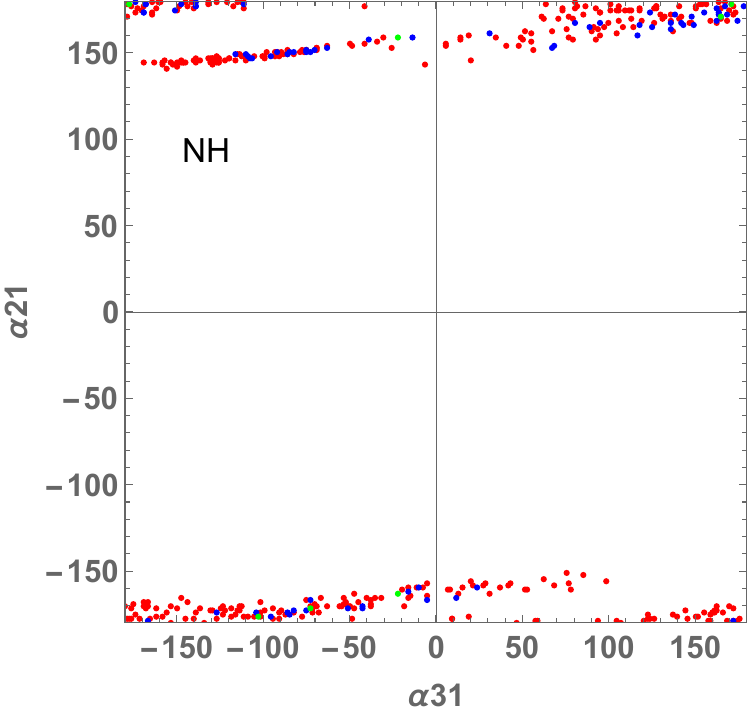} 
\caption{The values of Dirac CP phase and Majorana phases in NH case.  }   
\label{fig:3}\end{center}\end{figure}
%%%%%%%%%%%%%%%%%%%
%%%%%%%%%%%%%%%%%%%%%%%%%%%%%%%%%%%%%%%%%%%%%%%%%%%%%%%%%%%%%%%%%%%%%%%%%%%%%%%%%%%%

 %%%%%%%%%%%%%%%%%%%%%%%%%%%%%%%%%%%%%%%%%%%%%%%%%%%%%%%%%%%%%%%%%%%%%%%%%%%%%%%%%%%%
%%%%%%%%%%%%%%%%%%%
\begin{figure}[tb!]\begin{center}
\includegraphics[width=65mm]{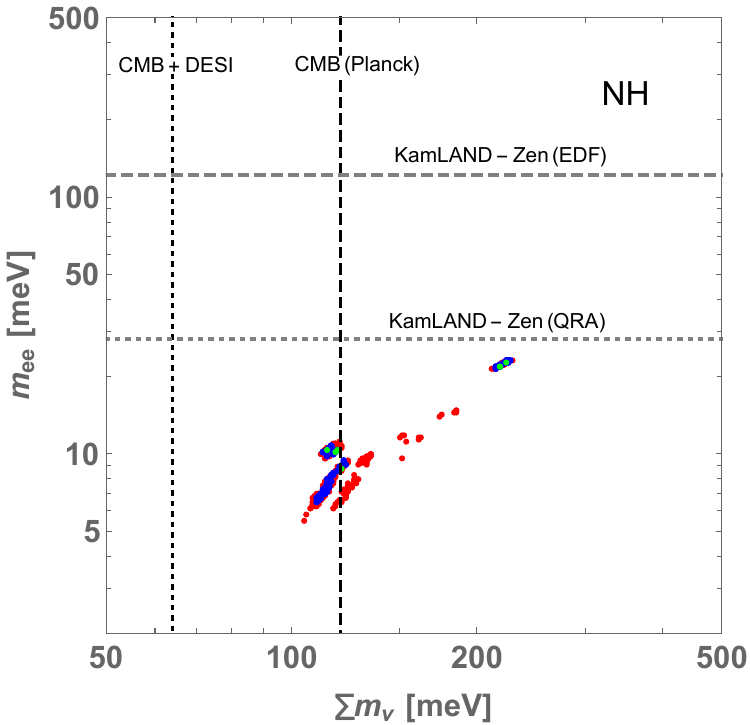} \
\includegraphics[width=65mm]{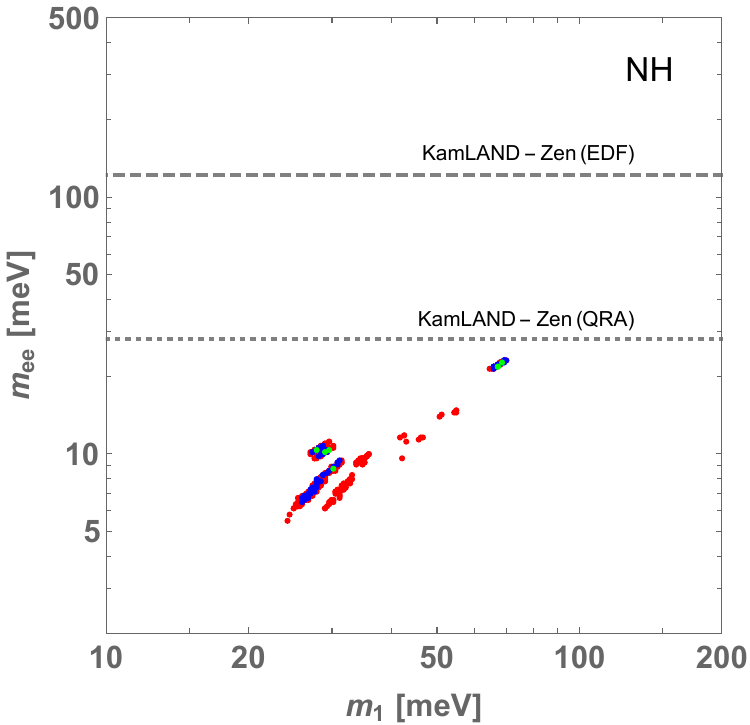} 
\caption{Neutrino mass related observables in NH case. Left: the obtained points on $\{\sum m_\nu, m_{ee} \}$ plane. Right: the points on $\{m_1, m_{ee} \}$ plane. 
The dashed and dotted vertical lines in the left panel indicate the upper bounds on $\sum m_\nu$ from CMB (Planck) data and combination of CMB and DESI data respectively.}   
\label{fig:4}\end{center}\end{figure}
%%%%%%%%%%%%%%%%%%%
%%%%%%%%%%%%%%%%%%%%%%%%%%%%%%%%%%%%%%%%%%%%%%%%%%%%%%%%%%%%%%%%%%%%%%%%%%%%%%%%%%%%

In Fig.~\ref{fig:3}, we show CP and Majorana phases obtained from the allowed parameters in NH case; the left and right panels respectively
 represent $\{\delta_{CP}, \alpha_{21} \}$ and $\{\alpha_{31}, \alpha_{21} \}$ planes.
We obtain a wide range of Dirac CP phase  where the region $[-130, -70]$ [deg] is not favored while there is no specific correlation between $\delta_{CP}$ and $\alpha_{21}$.
The value of Majorana phase $\alpha_{21}$ is restricted in region around $[140, 180]$ [deg] and $[-180, -150]$ [deg] while the value of $\alpha_{31}$ is not restricted.

In Fig.~\ref{fig:4}, the left panel shows the obtained points on $\{\sum m_\nu, m_{ee} \}$ plane while the right panel shows those on $\{m_1, m_{ee} \}$ plane.
The dashed and dotted vertical lines in the left panel indicate the upper bounds on $\sum m_\nu$ from CMB (Planck) data and combination of CMB and DESI data respectively.
We find that some points satisfy the CMB $\sum m_\nu$ constraint but all the points are in tension with the "CMB+DESI" limit.
The obtained values of $m_{ee}$ are below the current KamLAND-Zen constraint and some points could be tested in future experiments.

\subsubsection{IH case}

 %%%%%%%%%%%%%%%%%%%%%%%%%%%%%%%%%%%%%%%%%%%%%%%%%%%%%%%%%%%%%%%%%%%%%%%%%%%%%%%%%%%%
%%%%%%%%%%%%%%%%%%%
\begin{figure}[tb!]\begin{center}
\includegraphics[width=50mm]{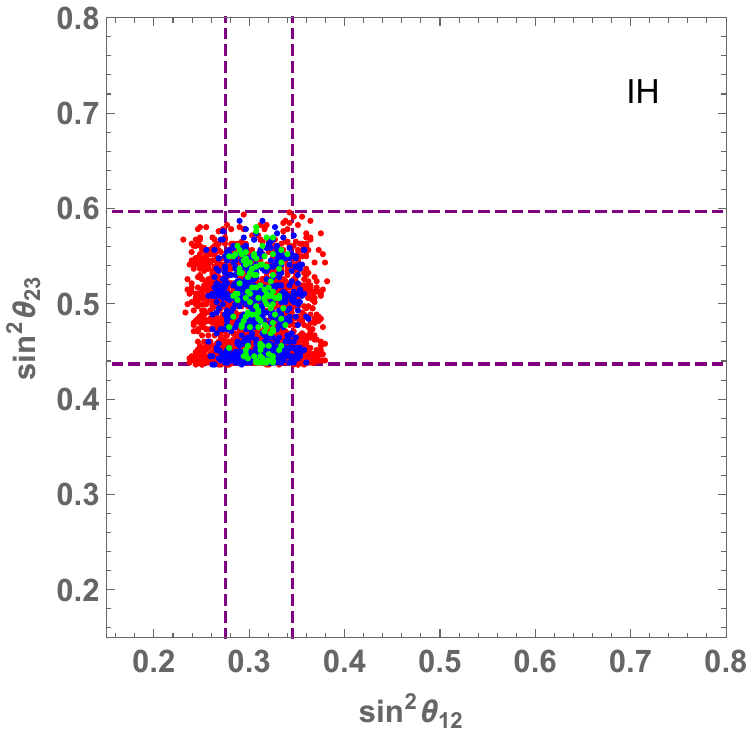} \
\includegraphics[width=53mm]{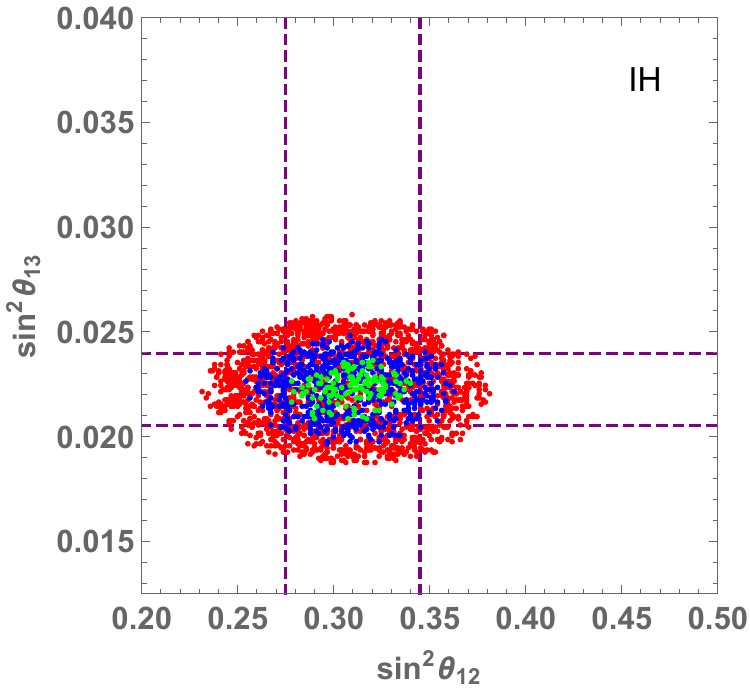} \
\includegraphics[width=53mm]{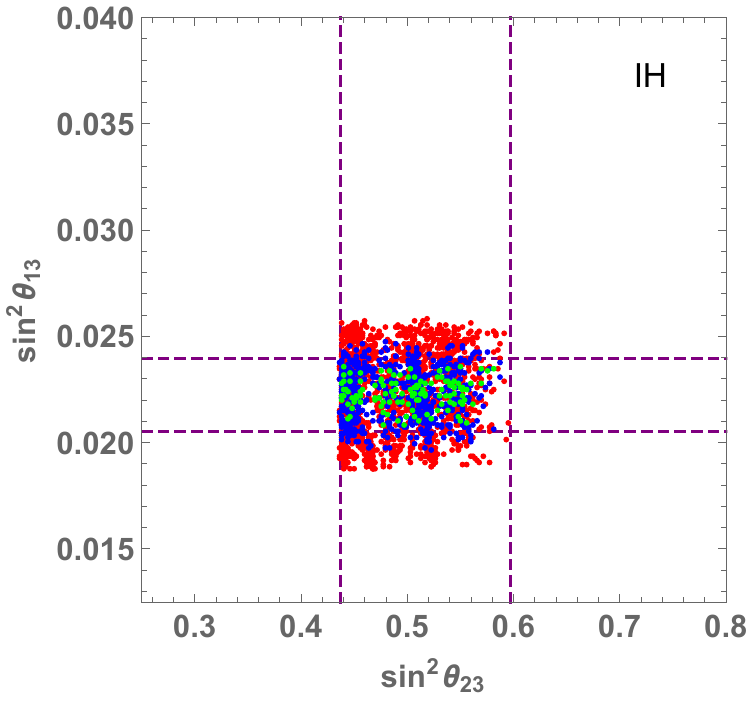} 
\caption{The neutrino mixing angles obtained from the allowed parameter points in IH case where the region between purple dashed line correspond to $3 \sigma$ range of observables.  }   
\label{fig:5}\end{center}\end{figure}
%%%%%%%%%%%%%%%%%%%
%%%%%%%%%%%%%%%%%%%%%%%%%%%%%%%%%%%%%%%%%%%%%%%%%%%%%%%%%%%%%%%%%%%%%%%%%%%%%%%%%%%%

In Fig.~\ref{fig:5}, we show the neutrino mixing angles obtained from the allowed parameter points in IH case where the regions between purple dashed lines correspond to $3 \sigma$ range of the observed values.
We find that most of the region in 3$\sigma$ range is realized by our allowed points where the region $0.58 \lesssim \sin^2 \theta_{23}$ is slightly disfavored.

 %%%%%%%%%%%%%%%%%%%%%%%%%%%%%%%%%%%%%%%%%%%%%%%%%%%%%%%%%%%%%%%%%%%%%%%%%%%%%%%%%%%%
%%%%%%%%%%%%%%%%%%%
\begin{figure}[tb!]\begin{center}
\includegraphics[width=65mm]{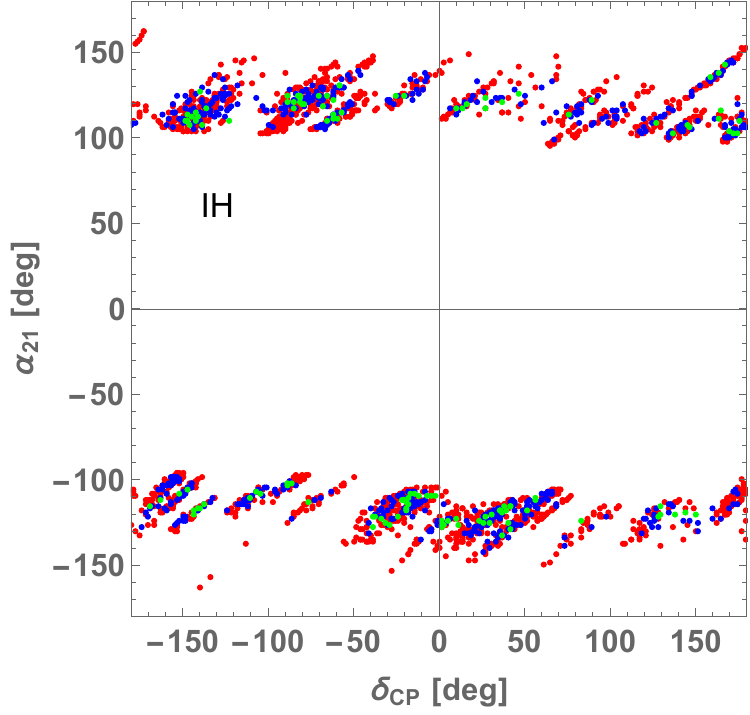} \
\includegraphics[width=65mm]{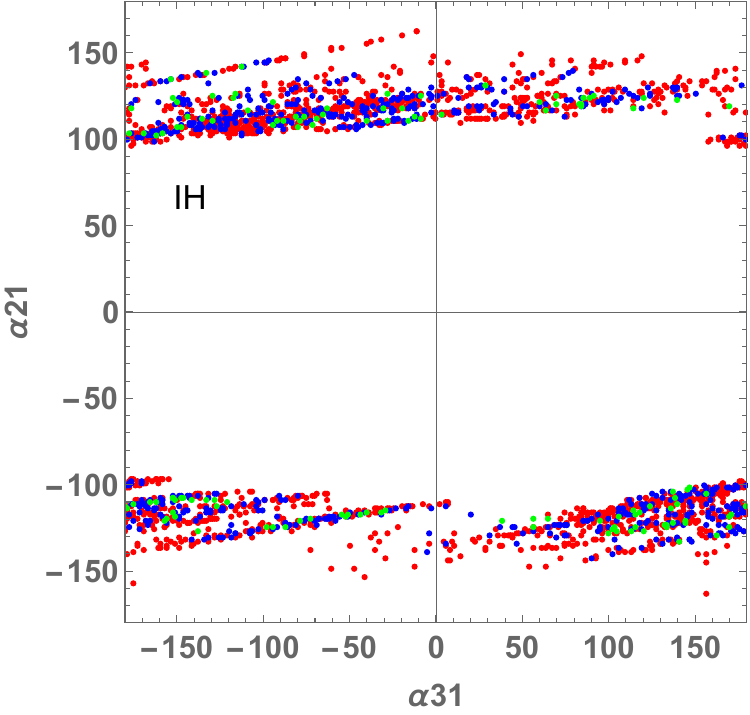} 
\caption{The values of Dirac CP phase and Majorana phase in IH case.  }   
\label{fig:6}\end{center}\end{figure}
%%%%%%%%%%%%%%%%%%%
%%%%%%%%%%%%%%%%%%%%%%%%%%%%%%%%%%%%%%%%%%%%%%%%%%%%%%%%%%%%%%%%%%%%%%%%%%%%%%%%%%%%

In Fig.~\ref{fig:6}, we show CP and Majorana phases obtained from the allowed parameters in IH case; the left and right panels respectively
 represent $\{\delta_{CP}, \alpha_{21} \}$ and $\{\alpha_{31}, \alpha_{21} \}$ planes.
Both figures tend to be linearly correlated.
We find allowed regions for the Dirac CP phase and $\alpha_{31}$ are not restricted.
On the other hand,
%and there is almost no correlation between $\delta_{21}$ and $\alpha_{31}$.
the value of Majorana phase $\alpha_{21}$ is favored in region around $[100, 150]$ [deg] and $[-150, -100]$ [deg]. 
%while the value of $\alpha_{31}$ is not restricted.

 %%%%%%%%%%%%%%%%%%%%%%%%%%%%%%%%%%%%%%%%%%%%%%%%%%%%%%%%%%%%%%%%%%%%%%%%%%%%%%%%%%%%
%%%%%%%%%%%%%%%%%%%
\begin{figure}[tb!]\begin{center}
\includegraphics[width=65mm]{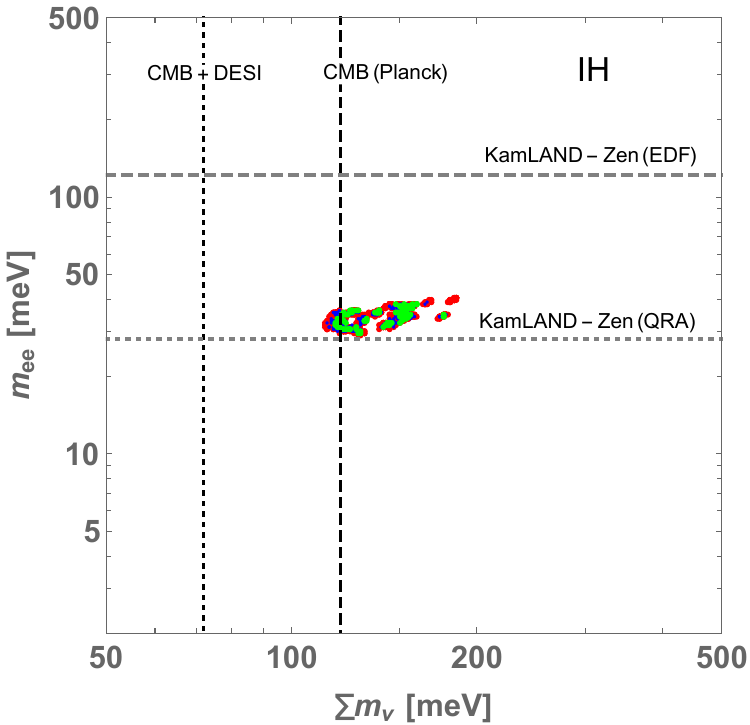} \
\includegraphics[width=65mm]{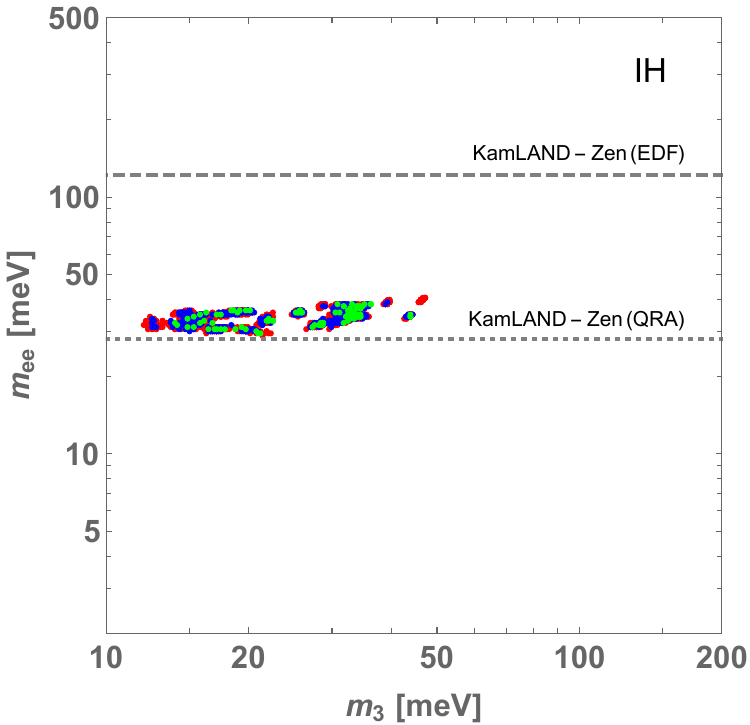} 
\caption{Neutrino mass related observables in IH case. Left: the obtained points on $\{\sum m_\nu, m_{ee} \}$ plane. Right: the points on $\{m_3, m_{ee} \}$ plane.
The horizontal and vertical lines are the same as in the Fig.~\ref{fig:4}.  }   
\label{fig:7}\end{center}\end{figure}
%%%%%%%%%%%%%%%%%%%
%%%%%%%%%%%%%%%%%%%%%%%%%%%%%%%%%%%%%%%%%%%%%%%%%%%%%%%%%%%%%%%%%%%%%%%%%%%%%%%%%%%%

In Fig.~\ref{fig:7}, the left panel shows the allowed points on $\{\sum m_\nu, m_{ee} \}$ plane while the right panel shows those on $\{m_3, m_{ee} \}$ plane for IH case.
The dashed and dotted vertical lines are the same as in Fig.~\ref{fig:4}.
We find that the range is narrower than the case of NH where $\sum m_\nu$ is within around $[115, 180]$ [meV] and $m_{ee}$ is within around $[28, 41]$ [meV].  
Some points satisfy the CMB $\sum m_\nu$ constraint but all the points are in tension with the "CMB+DESI" limit as in the NH case.
The obtained values of $m_{ee}$ are in the range of the current KamLAND-Zen constraint and we expect the points can be tested in near future.

\section{Summary and discussion}

We have studied a radiative seesaw model with leptoquarks under non-holomorphic modular $A_4$ symmetry.
The leptons and quarks are assigned to be $A_4$ triplets with zero modular weight except for right-handed up-type quarks that are singlets with weight $-6$, as our simplest choice to fit the observed data.
 The quark and charged lepton matrices have been formulated requiring invariance under the modular symmetry. 
 The neutrino masses are generated at one-loop level where the structure of relevant Yukawa couplings is also restricted by the modular symmetry.

We have carried out numerical analysis scanning free parameters in the model to fit observed data for both quark and lepton sectors.
In fitting quark sector, we have found that Im$[\tau] \sim 2.3-2.4$ is required and $| {\rm Re}[\tau]|$ is also preferred to be below $0.4$, 
and obtained allowed points of $\tau$ have been used in analyzing lepton sector.
In lepton sector, we have fitted observed values such as charged lepton masses, neutrino mixing angles and mass squared differences 
by scanning free parameters relevant to leptons. 
As a result we have found the model can fit neutrino observables for both NH and IH cases. 

For NH case, neutrino mixing angles are bit restricted where $\sin \theta_{12}$ value tends to be larger as $0.3 \lesssim \sin^2 \theta_{12}$ for $0.57 \lesssim \sin^2 \theta_{23}$ and $0.32 \lesssim \sin^2 \theta_{12}$ for entire range of $\sin^2 \theta_{23}$ within $3 \sigma$ level.
Also Dirac CP phase disfavors the range of $[-130, -70]$ [deg] and no specific correlation has been found between $\delta_{CP}$ and $\alpha_{21(31)}$.
The value of Majorana phase $\alpha_{21}$ is restricted in region around $[140, 180]$ [deg] and $[-180, -150]$ [deg] while the value of $\alpha_{31}$ is not restricted.
The value of $m_{ee}$ is below the current KamLAND-Zen constraint and some points would be tested in future.
We have found that some points satisfy the CMB constraint on $\sum m_\nu$ but are in tension with the CMB+DESI constraint.

For IH case, neutrino mixing angles span almost entire regions within $3 \sigma$ level.
The Dirac CP phase is not restricted 
while it tends to be linearly correlated among the phases.
%and there is almost no correlation among phases.
The value of Majorana phase $\alpha_{21}$ is favored in region around $[100, 150]$ [deg] and $[-150, -100]$ [deg] while the value of $\alpha_{31}$ is not restricted.
We have found that the obtained region of $\{\sum m_\nu, m_{ee}\}$ is narrower than the case of NH where $\sum m_\nu$ is within around $[115, 180]$ [meV] and $m_{ee}$ is within around $[28, 41]$ [meV].  
Some points satisfy the CMB $\sum m_\nu$ constraint but all the points are in tension with the "CMB+DESI" limit as in the NH case.
The values of $m_{ee}$ are in the range of the current KamLAND-Zen constraint and we expect the points can be tested in near future.

In this work, we have focused on the quark and lepton masses including neutrino mass generation, but it is interesting to consider other phenomenology such as 
collider physics, lepton flavor violation and meson decays; we can easily avoid phenomenological constraints related to them by making leptoquark masses heavier without change of our predictions. 
We expect modular flavor symmetry would restrict behavior of these phenomena compared with the models without the symmetry that would provide specific upper limit on leptoquark masses.
We left the detailed investigations of them as upcoming works.

%\newpage
%%%%%%%%%%%%%%%%%%%%%%%%%%%%%%%%%%%
\section*{Acknowledgments}
The work was supported by the Fundamental Research Funds for the Central Universities (T.~N.).
%%%%%%%%%%%%%%%%%%%%%%%%%%%%%%%%%%%
%%%%%%%%%%%%%%%%%%%%%%%%%%%%%%%%%%%

\vspace{-3mm}

\bibliography{NonHoloMa4_LQ.bib}

\end{document}